\documentclass{article}
\usepackage{url}
\usepackage{authblk}
\usepackage{listings}
\usepackage{blindtext}
\usepackage{booktabs}
\usepackage{cite}
\usepackage{float}
\usepackage{tcolorbox}
\usepackage{longtable}
\usepackage{algorithm}
\usepackage{algorithmic}
\usepackage[utf8]{inputenc}
\usepackage[T1]{fontenc}
\usepackage{tcolorbox}
\usepackage{booktabs}
\usepackage{array}
\usepackage[colorlinks=True]{hyperref}
\usepackage[labelfont=bf]{caption}
\usepackage[width=18.00cm, height=26.00cm,margin=1in]{geometry}
\usepackage{graphicx} 
\usepackage[document]{ragged2e}
\title{\textbf{COS-Mix: Cosine Similarity and Distance Fusion for Improved Information Retrieval}}
\author{Kush Juvekar$^1$ and Anupam Purwar$^2$$^*$\\$^1$ \url{https://gihub.com/Koosh0610}, Ahmedabad, India \\$^2$ \url{https://anupam-purwar.github.io/page/}, Delhi, India}
\affil[1,2]{Both authors contributed equally to this research.}
\affil[*]{Corresponding author: Anupam Purwar, anupam.aiml@gmail.com}

\date{June 2024}

\begin{document}
\vspace{4em}
\maketitle

\begin{abstract}
\addcontentsline{toc}{chapter}{Abstract}
This study proposes a novel hybrid retrieval strategy for Retrieval-Augmented Generation (RAG) that integrates cosine similarity and cosine distance measures to improve retrieval performance, particularly for sparse data. The traditional cosine similarity measure is widely used to capture the similarity between vectors in high-dimensional spaces. However, it has been shown that this measure can yield arbitrary results in certain scenarios. To address this limitation, we incorporate cosine distance measures to provide a complementary perspective by quantifying the dissimilarity between vectors. Our approach is experimented on proprietary data, unlike recent publications that have used open-source datasets. The proposed method demonstrates enhanced retrieval performance and provides a more comprehensive understanding of the semantic relationships between documents or items. This hybrid strategy offers a promising solution for efficiently and accurately retrieving relevant information in knowledge-intensive applications, leveraging techniques such as BM25 (sparse) retrieval , vector (Dense) retrieval, and cosine distance based retrieval to facilitate efficent information retrieval. \\
\vspace{1em}
\end{abstract}

\section{Introduction}

{Large Language Models (LLMs) have emerged as transformative technologies with excellent performance on a variety of tasks. With the increasing size of LLMs, they can function as very effective knowledge warehouses \cite{petroni2019language}, with facts embedded within their parameters, and models can be improved further through fine-tuning on domain-specific knowledge. However fine-tuning is a difficult task with vast amounts of data \cite{10.5555/3495724.3496517}. A different method, first developed in open domain question answering systems \cite{chen-etal-2017-reading}, involves organizing vast amounts of text into smaller sections (paragraphs) and storing them in a distinct information retrieval system. This system retrieves relevant information, which is then provided to the LLM alongside the question for context.  Researchers have also attempted using keywords to augment information retrieval \cite{10.1145/3639856.3639916} with reported reduction in latency and cost of retrieval \cite{10.1145/3639856.3639916}. This approach simplifies the process of supplying a system with up-to-date knowledge in a specific domain, while also facilitating easy understanding of where the information comes from. In contrast, the inherent knowledge of LLMs is complex and challenging to trace back to its origin \cite{akyurek2022tracing}.\\
\vspace{1em}
Nevertheless, existing retrieval-augmented approaches also have flaws. Most practices in retrieval augmented generation or RAG, use vector similarity as semantic similarity, but it has been shown that cosine similarity of learned embeddings can yield arbitrary results \cite{10.1145/3589335.3651526}. In this study, we propose a hybrid retrieval strategy that integrates cosine similarity and cosine distance measures to enhance retrieval performance. Traditional cosine similarity measures have been widely utilized to capture the similarity between vectors in high-dimensional spaces. However, in scenarios where the similarity measure fails to adequately capture the semantic relationship between documents or items, cosine distance provides a complementary perspective by quantifying the dissimilarity between vectors. We demonstrate how this approach improves retrieval specifically for sparse data.\\
\vspace{1em}
Our proposed method towards RAG is experimented on proprietary data, unlike more recent publications \cite{gutiérrez2024hipporag,yan2024corrective,jeong2024adaptiverag} which have used open-source datasets like \textbf{QuALITY} \cite{pang-etal-2022-quality}, \textbf{MedQA} \cite{app11146421}, US SEC Filings etc. }

\begin{figure}
    \includegraphics[width=15cm]{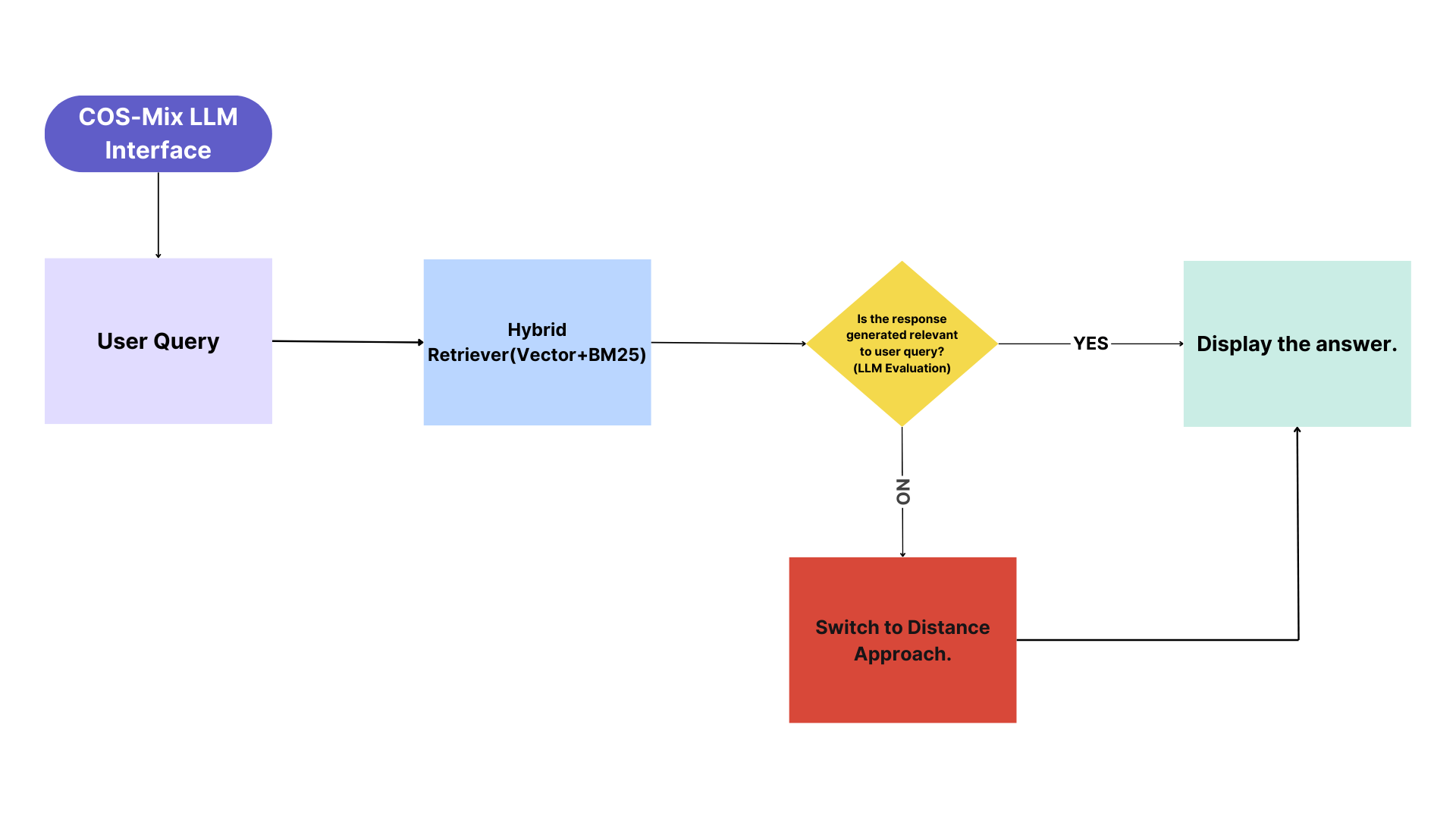}
    \caption{COS-Mix LLM Interface}
    \label{fig:1}
\end{figure}

\section{Methodology}
\subsection{Data Processing}
The HTML pages used in the study and experiments came from \url{https://i-venture.org}. A web crawler was set up to methodically obtain all HTML pages, using libraries like requests, re, urllib, BeautifulSoup, collections, and HTML parsing tools. Upon recovery, the HTML pages were converted to text files. Following that, a variety of data preparation techniques were used to remove unnecessary information such as headers and footers, ensuring that the focus remained on the primary textual content.
\subsection{Vectorization}
The analyzed text files were divided into manageable pieces and then transformed into embeddings using the OpenAI embedding model \textbf{text-embedding-ada-002}. The size of those manageable pieces were optimized so as to achieve higher score on metrics discussed in \ref{subsection: evaluation}. During this procedure, entities were extracted for use as metadata in question answering \cite{Aarsen_SpanMarker}. The generated embedding vectors were systematically stored for further analysis.

\subsection{Retrieval}
Leveraging the Language Model (LLM) capabilities, user queries were addressed by invoking the assistance of the pre-trained language model, \textbf{GPT-3.5-TURBO}. In instances where the LLM failed to provide satisfactory responses based on contextual information(similarity search), the system seamlessly transitioned to a distance-based approach for query resolution. For RAG, we use a hybrid retriever composed of BM25 retriever coupled with traditional vector retriever and then the retrieved chunks were reordered \cite{Liu2023LostIT}. BM25 is a widely used information retrieval technique that employs a probabilistic model to rank documents based on the frequency and distribution of query terms within them \cite{10.1561/1500000019}. The chunks retrieved using distance approach are re-ranked with the help flashrank library \cite{Damodaran_FlashRank_Lightest_and_2023}.\\
\vspace{1em}
To facilitate the transition between the similarity-based and distance-based approaches, a validation prompt was utilized. This prompt, designed to evaluate the adequacy of responses generated by the initial approach, ensured a seamless and effective switch between methodologies based on the user's query requirements. Following is the prompt employed:\\
\vspace{1em}
\begin{tcolorbox}[colback=blue!5!white, colframe=blue!75!black, title= Validating Prompt]
"You are an intelligent bot designed to assist users on an organization's\\ website by answering their queries. You'll be given a user's question and\\ an associated answer. Your task is to determine if the provided answer\\ effectively resolves the query. If the answer is unsatisfactory, return 0.\\
Query:  \{query\}\\
Answer: \{answer\}\\
Your Feedback:"
\end{tcolorbox}

\subsection{Evaluation}
We evaluated our augmented-RAG pipeline on a variety of metrics and compared it with two open source datasets \cite{10.1007/978-3-319-30671-1_58,wadden-etal-2020-fact}. The evolution of Natural Language Processing (NLP) has transitioned from classical metrics like \textbf{ROUGE}, and \textbf{METEOR}\cite{banerjee-lavie-2005-meteor} to more nuanced metrics \cite{DeepEval}. A common issue in evaluating RAG is the unavailability of ground truths. We use answers generated by GPT-4-TURBO as ground truths as it has been demonstrated that LLMs like \textbf{GPT-4-TURBO} are over 80\% in agreement with humans and hence can be used as a judge for evaluating a smaller model’s output \cite{Zheng2023JudgingLW}. The output generated has been evaluated on the following nuanced metrics alongside the classical methods: \\
\vspace{1em}
\textbf{Contextual Precision}: This metric measures your RAG pipeline's retriever by evaluating whether nodes in your retrieval context that are relevant to the given query are ranked higher than irrelevant ones.\\
\vspace{1.5em}
\textbf{Contextual Recall}: This metric measures the quality of the RAG pipeline's retriever by evaluating the extent to which the retrieval context aligns with the ground truth.\\
\vspace{1em}
\textbf{Contextual Relevancy}: This metric measures the quality of the RAG pipeline's retriever by evaluating the overall relevance of the information presented in your retrieval context for a given query.\\
\vspace{1em}
\textbf{Answer Relevancy}: The answer relevancy metric measures the quality of the RAG pipeline's generator by evaluating how relevant the actual output of LLM is compared to the provided query.\\
\vspace{1em}
\textbf{Faithfulness}: This metric measures the quality of your RAG pipeline's generator by evaluating whether the actual output factually aligns with the contents of your retrieval context.\\
\label{subsection: evaluation}

\section{RESULTS AND DISCUSSION}
\subsection{Limitations of RAG}
Recent findings \cite{guinet2024automated} show that optimal choice of retrieval method and LLM is task-dependent and choice of retrieval method often impacts performance more than scaling LLM size. While the hybrid retrieval performs better on the proprietary dataset across almost all metrics (Table \ref{Table1}), it still performs poorly on contextual relevancy across all datasets. On an average, more than half of the context retrieved is not relevant to the user query or that the information is sparse meaning only few chunks contain relevant information. Besides, many times there are instances where the LLM fails to answer user queries even when the information to generate the answer is in the retrieved context.

\vspace{1em}
\label{subsection : 3.1}

\begin{table}[H]
\vspace{-1em}
\centering
\begin{tabular}{@{}cccc@{}}
\toprule
\multicolumn{4}{c}{\textbf{CLASSICAL METRICS}}                                                      \\ \midrule
\textbf{METRICS}              & \textbf{NFCorpus} & \textbf{SciFact} & \textbf{Proprietary Dataset} \\
\textbf{Precision}            & 0.55              & 0.60             & \textbf{0.77}                \\
\textbf{Recall}               & 0.40              & 0.39             & \textbf{0.63}                \\
\textbf{F-Score}              & 0.44              & 0.44             & \textbf{0.68}                \\
\textbf{METEOR}               & 0.38              & 0.40             & \textbf{0.66}                \\ \midrule
\multicolumn{4}{c}{\textbf{NUANCED METRICS}}                                                        \\ \midrule
\textbf{METRICS}              & \textbf{NFCorpus} & \textbf{Scifact} & \textbf{Proprietary Data}    \\
\textbf{Contextual Recall}    & 0.51              & 0.46             & \textbf{0.86}                \\
\textbf{Contextual Precision} & 0.83              & 0.80             & \textbf{0.98}                \\
\textbf{Contextual Relevancy} & \textbf{0.40}     & \textbf{0.43}    & \textbf{0.60}                \\
\textbf{Answer Relevancy}     & 0.87              & 0.90             & \textbf{1}                   \\
\textbf{Faithfulness}         & 0.94              & 0.95             & 0.90                         \\ \bottomrule
\end{tabular}
\caption{ Comprehensive overview classical and nuanced metrics for Hybrid Retrieval. All values are reported on average.}
\label{Table1}
\end{table}

\subsection{Efficacy of Distance Approach}
As discussed in Section \ref{subsection : 3.1}, classical RAG based on hybrid retriever fails to answer many questions with sparse information even after varying the chunk size and top-k values. However, the usage of distance approach in information retrieval augments the classical RAG and LLM is able to respond to such questions with accurate answers every time, as shown in Table \ref{table: table2}. Thus, distance approach augmented RAG performs without the need to use a more advanced LLM with higher context window and thereby saving cost (All responses in Table \ref{table: table2} are generated using GPT-3.5-TURBO).

\begin{longtable}{@{}clc@{}}
\caption{Efficacy of Distance Approach: Answers and Response Time } \\
\toprule
\textbf{S.NO}        & \multicolumn{1}{c}{\textbf{User Query and Answers}}                                                                                                                                                                                                                                                                                                                                                                                                                                                                                                                                                                                                                                                                                          & \multicolumn{1}{l}{\textbf{Response Time (s)}} \\ \midrule
\endfirsthead
\toprule
\textbf{S.NO}        & \multicolumn{1}{c}{\textbf{User Query and Answers}}                                                                                                                                                                                                                                                                                                                                                                                                                                                                                                                                                                                                                                                                                          & \multicolumn{1}{l}{\textbf{Response Time (s)}} \\ \midrule
\endhead
\midrule \multicolumn{3}{r}{\textit{Continued on next page}} \\ \midrule
\endfoot
\bottomrule
\endlastfoot
\textbf{1.}          & \textbf{What is the TRICOR Batch 4 program?}                                                                                                                                                                                                                                                                                                                                                                                                                                                                                                                                                                                                                                                                                                  & \textbf{}                                      \\ \midrule
\textbf{}            & \begin{tabular}[c]{@{}l@{}}\textbf{Hybrid Retrieval} : I'm sorry, but the information provided in the context\\ does not contain details about the TRICOR Batch 4 program. If you have\\ any other questions or need information on a different topic, feel free to\\ ask!\end{tabular}                                                                                                                                                                                                                                                                                                                                                                           & 8.165937                                       \\ \midrule
\multicolumn{1}{l}{} & \begin{tabular}[c]{@{}l@{}}\textbf{Distance Approach} : The TRICOR Batch 4 program is a flagship\\ entrepreneurship program in collaboration with the Tribal Cooperative\\ Finance Corporation Ltd (TRICOR) under the Telangana Government's \\ Tribal Welfare Department.\end{tabular}                                                                                                                                                                                                                                                                                                                                                                               & \textbf{4.632708}                              \\ \midrule
\textbf{2.}          & \textbf{How does TRICOR support tribal entrepreneurs?}                                                                                                                                                                                                                                                                                                                                                                                                                                                                                                                                                                                                                                                                                          & \textbf{}                                      \\ \midrule
\textbf{}            & \begin{tabular}[c]{@{}l@{}}\textbf{Hybrid Retrieval} : I'm sorry, but the information provided in the context\\ does not contain details about how TRICOR supports tribal entrepreneurs. \\ If you have any other questions or need information on a different topic,\\ feel free to ask!\end{tabular}                                                                                                                                                                                                                                                                                                                                                             & 5.847167                                       \\ \midrule
\multicolumn{1}{l}{} & \begin{tabular}[c]{@{}l@{}}\textbf{Distance Approach} : TRICOR supports tribal entrepreneurs\\ through capacity-building sessions and other resources.\end{tabular}                                                                                                                                                                                                                                                                                                                                                                                                                                                                                                                                                                                & \textbf{3.824327}                              \\ \midrule
\textbf{3.}          & \textbf{What is Atal Catalysts @ ISB 2.0?}                                                                                                                                                                                                                                                                                                                                                                                                                                                                                                                                                                                                                                                                                                      & \textbf{}                                      \\ \midrule
\textbf{}            & \begin{tabular}[c]{@{}l@{}}\textbf{Hybrid Retrieval} : I'm sorry, but the information provided in the context\\ does not include details about the Atal Catalyst @ ISB. If you have any\\ other questions or need information on a different topic, feel free to ask!\end{tabular}                                                                                                                                                                                                                                                                                                                                                                                   & 7.982479                                       \\ \midrule
\multicolumn{1}{l}{} & \begin{tabular}[c]{@{}l@{}}\textbf{Distance Approach} : Atal Catalysts @ ISB 2.0 is a mentorship program in \\ partnership with Atal Innovation Mission (AIM), NITI Aayog.\end{tabular}                                                                                                                                                                                                                                                                                                                                                                                                                                                                                                                                                            & \textbf{4.338906}                              \\ \midrule
\textbf{4.}          & \textbf{Who is the CEO of DLabs?}                                                                                                                                                                                                                                                                                                                                                                                                                                                                                                                                                                                                                                                                                                               & \textbf{}                                      \\ \midrule
\textbf{}            & \begin{tabular}[c]{@{}l@{}}\textbf{Hybrid Retrieval} : The CEO of DLabs is not explicitly mentioned in\\ the provided excerpts. However, based on the information available, the\\ positions mentioned in the documents related to I-Venture @ ISB are:\\ \\ Assistant Program Manager at AIC ISB\\ Executive Admin \& Operations at AIC ISB\\ \\ If you have any specific questions regarding these positions or any other\\ information related to I-Venture @ ISB, feel free to ask!\end{tabular}                                                                                                                                                                                                            & 7.726001                                       \\ \midrule
\multicolumn{1}{l}{} & \begin{tabular}[c]{@{}l@{}}\textbf{Distance Approach} : Saumya Kumar is the CEO of DLabs at the Indian \\ School of Business (ISB).\end{tabular}                                                                                                                                                                                                                                                                                                                                                                                                                                                                                                                                                                & \textbf{4.143123}                              \\ \midrule
\textbf{5.}          & \textbf{Who is the assistant program manager at AIC ISB?}                                                                                                                                                                                                                                                                                                                                                                                                                                                                                                                                                                                                                                                                                        & \textbf{}                                      \\ \midrule
\textbf{}            & \begin{tabular}[c]{@{}l@{}}\textbf{Hybrid Retrieval} : The Assistant Program Manager at AIC ISB is \\ responsible for assisting the program manager in designing, promoting, \\ and managing the AIC ISB incubation/acceleration programs. Here are\\ some details about the Assistant Program Manager position at AIC ISB:\\ \\ Roles and Responsibilities:\\ \\ Assistant Program Managers in the new programs within AIC incubation,\\ acceleration, and corporate innovation verticals\\ Supporting end-to-end management of programs along with stakeholder\\ management\\ Engaging proactively with the startup ecosystem in India\\ Developing and managing a pipeline of mentors, industry experts, and\\ advisors for AIC across diverse industries\\ Working closely with ISB and its faculty members to leverage their subject\\ matter expertise..................................\end{tabular} & 9.692360                                       \\ \midrule
\multicolumn{1}{l}{} & \begin{tabular}[c]{@{}l@{}}\textbf{Distance Approach} : The assistant program manager at AIC ISB is\\ Saitejeswara Reddy.\end{tabular}                                                                                                                                                                                                                                                                                                                                                                                                                                                                                                                                                                          & \textbf{9.280980}                              \\ \bottomrule
\label{table: table2}
\end{longtable}

The above examples in Table \ref{table: table2}, demonstrate that  how distance approach can effectively answer questions with no compromise in response time. For example, the answer from hybrid retrieval in the 4\textsuperscript{th} question fails to answer who the CEO of D-Labs is but mentions that it has information about the Assistant Program Manager at AIC ISB. However upon asking that question we see that it doesn’t answer the question but just describes the role. On the other hand, using the distance approach we are able to  retrieve the most relevant context and then the LLM is able to generate the answer successfully. 

\section{Proposed Algorithm}
The proposed algorithm helps address the problem of time spent in information retrieval of sparse information during inference from the a large corpus of chunks $T$. It solves for this by identifying a priori all chunks which correspond to sparse information and then, creating a sub-set of these chunks such that $S \subseteq T$. Rest of the chunks form the set $R$ such that $R \subseteq T$ Thus, one does not have to sequentially search all the text chunks to identify the most relevant chunks when the hybrid retrieval fails in retrieving relevant context. In this case, the algorithm switches to the distance approach and searches in the set of sparse chunks ($S$) only, thereby reducing the time to retrieve compared to searching for all the chunks. \\

\newpage

Part 1: Create subset $S$ from corpus $T$::
\vspace{-1em}
\begin{algorithm}[H]
\begin{algorithmic}[1]
\REQUIRE Corpus $T$ consisting of chunks
\ENSURE Subset $S$ with sparse information chunks, subset $R$ with non-sparse information chunks
\STATE Initialize $S \leftarrow \emptyset$
\STATE Initialize $R \leftarrow \emptyset$
\FOR{each chunk $c \in T$}
    \STATE $isSparse \leftarrow$ \texttt{identifySparseInformation}($c$, $T$) \COMMENT{Check if information in $c$ occurs only in $c$}
    \IF{$isSparse$}
        \STATE $S \leftarrow S \cup \{c\}$ \COMMENT{Add chunk $c$ to subset $S$}
    \ELSE
        \STATE $R \leftarrow R \cup \{c\}$ \COMMENT{Add chunk $c$ to subset $R$}
    \ENDIF
\ENDFOR

\STATE \textbf{Function} \texttt{identifySparseInformation}($chunk$, $corpus$)
    \STATE Initialize $unique \leftarrow \texttt{True}$
    \FOR{each $otherChunk \in corpus$}
        \IF{$otherChunk \neq chunk$}
            \IF{information in $chunk$ is found in $otherChunk$}
                \STATE $unique \leftarrow \texttt{False}$
                \STATE \textbf{break}
            \ENDIF
        \ENDIF
    \ENDFOR
    \RETURN $unique$
\end{algorithmic}
\end{algorithm}

\vspace{-1em}
Part 2: During Inference::
\vspace{-1em}
\begin{algorithm}[H]
\begin{algorithmic}[1]
\REQUIRE User query $Q$, hybrid retriever $H$, large language model $LLM$, subsets of chunks $R$ and $S$ from total chunks $T$
\ENSURE Satisfactory answer to the user query
\STATE $chunks_R \leftarrow H.retrieve(R, Q)$ \COMMENT{Retrieve information from subset $R$}
\STATE $initial\_response \leftarrow LLM.generate(chunks_R)$ \COMMENT{Generate initial response}
\STATE $validation \leftarrow LLM.validate(Q, initial\_response)$ \COMMENT{Validate response}

\IF{$validation$ is satisfactory}
    \STATE Display $initial\_response$
\ELSE
    \STATE $chunks_S \leftarrow H.retrieve(S, Q)$ \COMMENT{Retrieve information from subset $S$}
    \STATE $final\_response \leftarrow LLM.generate(chunks_S)$ \COMMENT{Generate final response}
    \STATE Display $final\_response$
\ENDIF
\end{algorithmic}
\end{algorithm}
\vspace{0.5em}
By creating a small subset out of large corpus $T$, we avoid latency involved in calculating distance between embedding vectors during inference time. 
\vspace{1em}

\section{Conclusion}
There have been many methods to improve retrieval in recent publications \cite{10184013} making information retrieval one of the most active areas of research in Information Theory. Our experiments demonstrate that RAG over proprietary datasets is far more challenging than open source datasets prepared for evaluation as it fails to provide satisfactory answers for sparse information. However, enterprises cannot  afford to have an information retrieval solution which prevents LLMs from providing the correct response generation. This is where the distance approach can complement information retrieval of classical thereby allowing one to explore the full potential of LLMs in generating quality answers every time. We would like put forward our findings in the following key points:\\
\begin{itemize}
    \item The experiment demonstrates that Retrieval-Augmented Generation (RAG) on proprietary datasets poses significant challenges compared to open-source datasets.
    \item Classical RAG with hybrid retrieval often fail to provide satisfactory responses when dealing with sparse information, impacting the efficiency and reliability of information retrieval.
    \item A distance-based approach is proposed to complement hybrid retrieval, enhancing the overall retrieval process by focusing on sparse chunks of information.
    \item Experiments confirm that  distance approach does not compromise on quality of responses in scenarios where answers generated by LLM using context pulled using hybrid retrieval fail.
    \item The findings suggest that integrating the distance approach into classical RAG can unlock the full potential of Large Language Models (LLMs), ensuring accurate and efficient information retrieval for enterprise specific data.
\end{itemize}
\section*{Acknowledgement}
The authors thank  I-Venture at Indian School of Business for infrastructural support toward this work. Authors are extremely grateful to Prof. Bhagwan Chowdhry, Faculty Director (I-Venture at ISB) for his continued encouragement and support to carry out this research.

\bibliographystyle{IEEEtran}
\bibliography{sources}

\begin{thebibliography}{10}
\providecommand{\url}[1]{#1}
\csname url@samestyle\endcsname
\providecommand{\newblock}{\relax}
\providecommand{\bibinfo}[2]{#2}
\providecommand{\BIBentrySTDinterwordspacing}{\spaceskip=0pt\relax}
\providecommand{\BIBentryALTinterwordstretchfactor}{4}
\providecommand{\BIBentryALTinterwordspacing}{\spaceskip=\fontdimen2\font plus
\BIBentryALTinterwordstretchfactor\fontdimen3\font minus \fontdimen4\font\relax}
\providecommand{\BIBforeignlanguage}[2]{{%
\expandafter\ifx\csname l@#1\endcsname\relax
\typeout{** WARNING: IEEEtran.bst: No hyphenation pattern has been}%
\typeout{** loaded for the language `#1'. Using the pattern for}%
\typeout{** the default language instead.}%
\else
\language=\csname l@#1\endcsname
\fi
#2}}
\providecommand{\BIBdecl}{\relax}
\BIBdecl

\bibitem{petroni2019language}
F.~Petroni, T.~Rocktäschel, P.~Lewis, A.~Bakhtin, Y.~Wu, A.~H. Miller, and S.~Riedel, ``Language models as knowledge bases?'' 2019.

\bibitem{10.5555/3495724.3496517}
P.~Lewis, E.~Perez, A.~Piktus, F.~Petroni, V.~Karpukhin, N.~Goyal, H.~K\"{u}ttler, M.~Lewis, W.-t. Yih, T.~Rockt\"{a}schel, S.~Riedel, and D.~Kiela, ``Retrieval-augmented generation for knowledge-intensive nlp tasks,'' in \emph{Proceedings of the 34th International Conference on Neural Information Processing Systems}, ser. NIPS '20.\hskip 1em plus 0.5em minus 0.4em\relax Red Hook, NY, USA: Curran Associates Inc., 2020.

\bibitem{chen-etal-2017-reading}
\BIBentryALTinterwordspacing
D.~Chen, A.~Fisch, J.~Weston, and A.~Bordes, ``Reading {W}ikipedia to answer open-domain questions,'' in \emph{Proceedings of the 55th Annual Meeting of the Association for Computational Linguistics (Volume 1: Long Papers)}, R.~Barzilay and M.-Y. Kan, Eds.\hskip 1em plus 0.5em minus 0.4em\relax Vancouver, Canada: Association for Computational Linguistics, Jul. 2017, pp. 1870--1879. [Online]. Available: \url{https://aclanthology.org/P17-1171}
\BIBentrySTDinterwordspacing

\bibitem{10.1145/3639856.3639916}
\BIBentryALTinterwordspacing
A.~Purwar and R.~Sundar, ``Keyword augmented retrieval: Novel framework for information retrieval integrated with speech interface,'' in \emph{Proceedings of the Third International Conference on AI-ML Systems}, ser. AIMLSystems '23.\hskip 1em plus 0.5em minus 0.4em\relax New York, NY, USA: Association for Computing Machinery, 2024. [Online]. Available: \url{https://doi.org/10.1145/3639856.3639916}
\BIBentrySTDinterwordspacing

\bibitem{akyurek2022tracing}
\BIBentryALTinterwordspacing
E.~Akyurek, T.~Bolukbasi, F.~Liu, B.~Xiong, I.~Tenney, J.~Andreas, and K.~Guu, ``Towards tracing knowledge in language models back to the training data,'' in \emph{Findings of the Association for Computational Linguistics: EMNLP 2022}, Y.~Goldberg, Z.~Kozareva, and Y.~Zhang, Eds.\hskip 1em plus 0.5em minus 0.4em\relax Abu Dhabi, United Arab Emirates: Association for Computational Linguistics, Dec. 2022, pp. 2429--2446. [Online]. Available: \url{https://aclanthology.org/2022.findings-emnlp.180}
\BIBentrySTDinterwordspacing

\bibitem{10.1145/3589335.3651526}
\BIBentryALTinterwordspacing
H.~Steck, C.~Ekanadham, and N.~Kallus, ``Is cosine-similarity of embeddings really about similarity?'' in \emph{Companion Proceedings of the ACM on Web Conference 2024}, ser. WWW '24.\hskip 1em plus 0.5em minus 0.4em\relax New York, NY, USA: Association for Computing Machinery, 2024, p. 887–890. [Online]. Available: \url{https://doi.org/10.1145/3589335.3651526}
\BIBentrySTDinterwordspacing

\bibitem{gutiérrez2024hipporag}
\BIBentryALTinterwordspacing
B.~J. Gutiérrez, Y.~Shu, Y.~Gu, M.~Yasunaga, and Y.~Su, ``Hipporag: Neurobiologically inspired long-term memory for large language models,'' 2024. [Online]. Available: \url{https://arxiv.org/abs/2405.14831}
\BIBentrySTDinterwordspacing

\bibitem{yan2024corrective}
\BIBentryALTinterwordspacing
S.-Q. Yan, J.-C. Gu, Y.~Zhu, and Z.-H. Ling, ``Corrective retrieval augmented generation,'' 2024. [Online]. Available: \url{https://arxiv.org/abs/2401.15884}
\BIBentrySTDinterwordspacing

\bibitem{jeong2024adaptiverag}
\BIBentryALTinterwordspacing
S.~Jeong, J.~Baek, S.~Cho, S.~J. Hwang, and J.~C. Park, ``Adaptive-rag: Learning to adapt retrieval-augmented large language models through question complexity,'' 2024. [Online]. Available: \url{https://arxiv.org/abs/2403.14403}
\BIBentrySTDinterwordspacing

\bibitem{pang-etal-2022-quality}
\BIBentryALTinterwordspacing
R.~Y. Pang, A.~Parrish, N.~Joshi, N.~Nangia, J.~Phang, A.~Chen, V.~Padmakumar, J.~Ma, J.~Thompson, H.~He, and S.~Bowman, ``{Q}u{ALITY}: Question answering with long input texts, yes!'' in \emph{Proceedings of the 2022 Conference of the North American Chapter of the Association for Computational Linguistics: Human Language Technologies}, M.~Carpuat, M.-C. de~Marneffe, and I.~V. Meza~Ruiz, Eds.\hskip 1em plus 0.5em minus 0.4em\relax Seattle, United States: Association for Computational Linguistics, Jul. 2022, pp. 5336--5358. [Online]. Available: \url{https://aclanthology.org/2022.naacl-main.391}
\BIBentrySTDinterwordspacing

\bibitem{app11146421}
\BIBentryALTinterwordspacing
D.~Jin, E.~Pan, N.~Oufattole, W.-H. Weng, H.~Fang, and P.~Szolovits, ``What disease does this patient have? a large-scale open domain question answering dataset from medical exams,'' \emph{Applied Sciences}, vol.~11, no.~14, 2021. [Online]. Available: \url{https://www.mdpi.com/2076-3417/11/14/6421}
\BIBentrySTDinterwordspacing

\bibitem{Aarsen_SpanMarker}
\BIBentryALTinterwordspacing
T.~Aarsen, ``Spanmarker.'' [Online]. Available: \url{https://github.com/tomaarsen/SpanMarkerNER}
\BIBentrySTDinterwordspacing

\bibitem{Liu2023LostIT}
\BIBentryALTinterwordspacing
N.~F. Liu, K.~Lin, J.~Hewitt, A.~Paranjape, M.~Bevilacqua, F.~Petroni, and P.~Liang, ``Lost in the middle: How language models use long contexts,'' \emph{Transactions of the Association for Computational Linguistics}, vol.~12, pp. 157--173, 2023. [Online]. Available: \url{https://api.semanticscholar.org/CorpusID:259360665}
\BIBentrySTDinterwordspacing

\bibitem{10.1561/1500000019}
\BIBentryALTinterwordspacing
S.~Robertson and H.~Zaragoza, ``The probabilistic relevance framework: Bm25 and beyond,'' \emph{Found. Trends Inf. Retr.}, vol.~3, no.~4, p. 333–389, apr 2009. [Online]. Available: \url{https://doi.org/10.1561/1500000019}
\BIBentrySTDinterwordspacing

\bibitem{Damodaran_FlashRank_Lightest_and_2023}
\BIBentryALTinterwordspacing
P.~Damodaran, ``{FlashRank, Lightest and Fastest 2nd Stage Reranker for search pipelines.}'' Dec. 2023. [Online]. Available: \url{https://github.com/PrithivirajDamodaran/FlashRank}
\BIBentrySTDinterwordspacing

\bibitem{10.1007/978-3-319-30671-1_58}
\BIBentryALTinterwordspacing
V.~Boteva, D.~Gholipour, A.~Sokolov, and S.~Riezler, ``A full-text learning to rank dataset for medical information retrieval,'' in \emph{Advances in Information Retrieval}, N.~Ferro, F.~Crestani, M.-F. Moens, J.~Mothe, F.~Silvestri, G.~M. Di~Nunzio, C.~Hauff, and G.~Silvello, Eds.\hskip 1em plus 0.5em minus 0.4em\relax Cham: Springer International Publishing, 2016, pp. 716--722. [Online]. Available: \url{https://link.springer.com/chapter/10.1007/978-3-319-30671-1_58}
\BIBentrySTDinterwordspacing

\bibitem{wadden-etal-2020-fact}
\BIBentryALTinterwordspacing
D.~Wadden, S.~Lin, K.~Lo, L.~L. Wang, M.~van Zuylen, A.~Cohan, and H.~Hajishirzi, ``Fact or fiction: Verifying scientific claims,'' in \emph{Proceedings of the 2020 Conference on Empirical Methods in Natural Language Processing (EMNLP)}, B.~Webber, T.~Cohn, Y.~He, and Y.~Liu, Eds.\hskip 1em plus 0.5em minus 0.4em\relax Online: Association for Computational Linguistics, Nov. 2020, pp. 7534--7550. [Online]. Available: \url{https://aclanthology.org/2020.emnlp-main.609}
\BIBentrySTDinterwordspacing

\bibitem{banerjee-lavie-2005-meteor}
\BIBentryALTinterwordspacing
S.~Banerjee and A.~Lavie, ``{METEOR}: An automatic metric for {MT} evaluation with improved correlation with human judgments,'' in \emph{Proceedings of the {ACL} Workshop on Intrinsic and Extrinsic Evaluation Measures for Machine Translation and/or Summarization}, J.~Goldstein, A.~Lavie, C.-Y. Lin, and C.~Voss, Eds.\hskip 1em plus 0.5em minus 0.4em\relax Ann Arbor, Michigan: Association for Computational Linguistics, Jun. 2005, pp. 65--72. [Online]. Available: \url{https://aclanthology.org/W05-0909}
\BIBentrySTDinterwordspacing

\bibitem{DeepEval}
C.~AI, ``Deepeval: The llm evaluation framework,'' \url{https://github.com/confident-ai/deepeval}, 2024.

\bibitem{Zheng2023JudgingLW}
\BIBentryALTinterwordspacing
L.~Zheng, W.-L. Chiang, Y.~Sheng, S.~Zhuang, Z.~Wu, Y.~Zhuang, Z.~Lin, Z.~Li, D.~Li, E.~P. Xing, H.~Zhang, J.~Gonzalez, and I.~Stoica, ``Judging llm-as-a-judge with mt-bench and chatbot arena,'' \emph{ArXiv}, vol. abs/2306.05685, 2023. [Online]. Available: \url{https://api.semanticscholar.org/CorpusID:259129398}
\BIBentrySTDinterwordspacing

\bibitem{guinet2024automated}
\BIBentryALTinterwordspacing
G.~Guinet, B.~Omidvar-Tehrani, A.~Deoras, and L.~Callot, ``Automated evaluation of retrieval-augmented language models with task-specific exam generation,'' 2024. [Online]. Available: \url{https://arxiv.org/abs/2405.13622}
\BIBentrySTDinterwordspacing

\bibitem{10184013}
\BIBentryALTinterwordspacing
K.~A. Hambarde and H.~Proença, ``Information retrieval: Recent advances and beyond,'' \emph{IEEE Access}, vol.~11, pp. 76\,581--76\,604, 2023. [Online]. Available: \url{https://ieeexplore.ieee.org/document/10184013}
\BIBentrySTDinterwordspacing

\end{thebibliography}
\end{document}